\documentclass[preprintnumbers,10pt,nofootinbib]{revtex4}

\usepackage{amsmath,latexsym,amssymb,amsfonts}
\usepackage[dvips]{graphicx,color}
\usepackage{bm}

\begin{document}

\preprint{YITP-17-39}

\title{\textbf{Thermal activation of thin-shells
in anti-de Sitter black hole spacetime} }

\author{
\textsc{Pisin Chen$^{a,b,c,d}$}\footnote{{\tt pisinchen{}@{}phys.ntu.edu.tw}},
\textsc{Guillem Dom\`enech$^{e,f}$}\footnote{{\tt guillem.domenech{}@{}yukawa.kyoto-u.ac.jp}},
\textsc{Misao Sasaki$^{e,f}$}\footnote{{\tt misao{}@{}yukawa.kyoto-u.ac.jp}} 
and
\textsc{Dong-han Yeom$^{a}$}\footnote{{\tt innocent.yeom{}@{}gmail.com}}
}

\affiliation{
$^{a}$\small{Leung Center for Cosmology and Particle Astrophysics, National Taiwan University, Taipei 10617, Taiwan}\\
$^{b}$\small{Department of Physics, National Taiwan University, Taipei 10617, Taiwan}\\
$^{c}$\small{Graduate Institute of Astrophysics, National Taiwan University, Taipei 10617, Taiwan}\\
$^{d}$\small{Kavli Institute for Particle Astrophysics and Cosmology,
SLAC National Accelerator Laboratory, Stanford University, Stanford, California 94305, USA}\\
$^{e}$\small{Center for Gravitational Physics, Yukawa Institute for Theoretical Physics, Kyoto University, Kyoto 606-8502, Japan}\\
$^{f}$\small{International Research Unit of Advanced Future Studies, Kyoto University, Kyoto 606-8502, Japan}}

\begin{abstract}
We investigate thermal activation of thin-shells around anti-de Sitter black holes.
Under the thin-shell approximation, we extensively study the parameter 
region that allows a bubble nucleation bounded by a thin-shell
out of a thermal bath.
We show that in general if one fixes the temperature outside the shell, 
one needs to consider the presence of a conical deficit inside the shell 
in the Euclidean manifold, due to the lack of solutions with a smooth
manifold. We show that for a given set of theoretical parameters, i.e., 
vacuum and shell energy density, there is a finite range of black hole 
masses that allow this transition. Most interestingly, one of them describes
the complete evaporation of the initial black hole.
\end{abstract}

\maketitle

\newpage

\tableofcontents


\section{Introduction}

Many exciting theoretical results have been obtained through studies of
quantum effects around black holes. Among them the most intriguing one
is the discovery of the Hawking radiation \cite{Hawking:1974sw},
that is, a black hole is semi-classically a thermodynamic object having 
a finite temperature.
This, however, led to the so-called information loss paradox \cite{Hawking:1976ra}.
A possible resolution within the semi-classical approach is the black hole 
complementarity \cite{Susskind:1993if}. However, there are arguments 
against it \cite{Yeom:2008qw,Almheiri:2012rt}. 
This means that one should departure from the semi-classical approximation 
and give up either unitarity or semi-classical gravity 
(for alternative interpretations, see \cite{Chen:2014jwq}). 

Here, before asking these fundamental questions of quantum gravity,
we study black holes in the semi-classical approximation more carefully, 
in the hope that a better understanding of the nature of semi-classical black holes 
may give us more insight into the nature of fully quantum black holes.
More specifically, we consider the role of \textit{non-perturbative} effects, 
i.e., quantum transitions that involve a change in the geometry of spacetime, 
around black holes \cite{Sasaki:2014spa}.

In the path integral formulation, quantum and/or thermal transitions 
are described by the sum over all possible configurations that satisfy
appropriate boundary conditions. 
In the Euclidean path integral approach, both quantum and thermal transitions
are treated on the same footing. 
In this formulation, the largest contribution comes from stationary points of 
the action, i.e., classical solutions to the Euclidean equations of motion. 
These configurations are called \textit{instantons}.

In the absence of gravity, a thermal transition
is dominated by $O(3)$ symmetric instantons that satisfy periodicity in 
the Euclidean time, say $\beta$, which corresponds to the temperature $T=1/\beta$
of the system~\cite{Linde:1993xx}. In the high temperature limit,
$\beta\to0$, only static solutions are allowed, i.e.,
they become translational invariant in the Euclidean time 
direction \cite{Garriga:1994ut}. For generic temperature, there may be plural
instanton solutions, which may or may not be static.
On the other hand, in the zero temperature limit, $\beta\to\infty$,
only a pure quantum vacuum transition is possible, and it is 
dominated by instantons with an enhanced symmetry $O(4)$.

We assume these properties are still true in the presence of 
gravity.\footnote{There is an issue about the meaning of the temperature
	in systems with gravity particularly in the case where there is 
	a positive vacuum energy, i.e., in de Sitter background. If the initial state
	is a de Sitter vacuum with its curvature radius $H^{-1}$, its Euclidean
	version has a periodicity of $2\pi/H$ which corresponds to the
	temperature $T=H/2\pi$. This is known as the Gibbons-Hawking temperature.
	Thus pure quantum fluctuations in de Sitter space may be interpreted as thermal 
	ones. However, since instantons describing transitions to states with lower 
	vacuum energy are $O(4)$ symmetric in contrast to the case of thermal instantons
	in flat space which are only $O(3)$ symmetric, it is not clear at all if this
	thermal interpretation of de Sitter space can be justified.}
Although there is no rigorous proof, there exist a few reasonable
arguments and pieces of evidence supporting this assumption.
See e.g., \cite{Tanaka:1992zw, Gen:1999gi}.
Namely, at the semi-classical level, it is assumed that the path integral 
may be performed over all possible geometries with the Euclidean signature,
and it will be dominated by classical solutions to the Euclidean 
field equations \cite{Hartle:1983ai}, which in particular may 
include non-trivial changes in the geometry.

In this paper, we consider thermal transitions of an anti-de Sitter (AdS) 
spacetime with a black hole, i.e., a black hole spacetime with a negative 
vacuum energy. The reason for considering such a spacetime is that 
the negative vacuum energy warps the spacetime in such a way that it 
effectively makes the space a bounded box so that canonical 
thermal equilibrium may be realized, unlike the case of asymptotic flat 
spacetimes where there exists no canonical equilibrium.

Thermodynamics for a system with a black hole in thermal equilibrium with 
radiation was studied in a pioneering work by Hawking and Page \cite{Hawking:1982dh}.
They found that for a given temperature above a critical temperature 
there are two possible black hole solutions. The smaller black hole is 
unstable and can either tunnel to a bigger stable one or decay into radiation. 
Below the critical temperature no black hole solution is possible; 
this is known as the Hawking-Page phase transition. Interestingly, 
invoking the AdS/CFT correspondence, Witten showed that the Hawking-Page 
phase transition from a black hole to pure AdS corresponds to a 
deconfinement/confinement thermal phase transition in gauge theory \cite{Witten:1998zw}.

In our case, among all possible end states to which transitions
may occur, we focus on those having a static $O(3)$ symmetric thin-shell
where the value of the vacuum energy inside the shell 
is different from the initial vacuum energy.
This may be regarded as a simplified toy model for a scalar field 
theory with two potential minima; one lower than the other.
Hence the transitions we consider may be regarded as the
formation of spherically symmetric bubbles with lower vacuum energy
by thermal activation, i.e., analogous to spharelons. 
A similar situation was studied previously by Garriga and 
Megevand~\cite{Garriga:2004nm} in de Sitter space instead of AdS space
and transitions {\it to\/} spacetimes with a black hole instead of those
{\it from\/} such spacetimes. 

Our initial spacetime may be considered either as a system in canonical
equilibrium with temperature given by that of the initial black hole or 
the one in micro-canonical equilibrium with a fixed mass at infinity.
In both cases, we assume that the effects of thermal bath outside the
black hole may be ignored. 

In the case of canonical equilibrium, possible final states
will be either with or without a black hole. Let $\beta$ be
the periodicity corresponding to the initial state.
If the final state contains a black hole, its temperature will
be different from the initial black hole temperature.
Then the corresponding instanton will have a conical singularity
at the horizon, since the periodicity of the final black hole required by
its regularity will not match $\beta$. 
But since the time scale for the system to reach thermal equilibrium would 
be extremely long for a large, semi-classical black hole, it is
physically reasonable to assume that the temperature of the final state 
is the same as the initial temperature right after the transition. 
Therefore, we assume such instantons with a cusp do contribute
to the transition. If the final state has no black hole, the instanton will
be a pure static AdS instanton with the Euclidean time periodically
identified with the period $\beta$. This is an interesting case which
describes a complete evaporation of the initial black hole by thermal transition.
We find that this is possible for a finite range of the parameter space.
As one would naturally suspect, this case is found to be closely related 
to the Hawking-Page transition. We find the complete evaporation occurs
in the regime of small black hole masses, and the transition probability 
is very similar to that of the Hawking-Page transition.

In the case of microcanonical equilibrium, we fix the total mass of the
system to be $M_{+}$. In this case, the instanton will have a cusp
because the initial and final black hole masses are definitely different.
It then turns out that the probability of tunneling is essentially the
same as the canonical case if we identify the temperature with the 
black hole temperature.

This paper is organized as follows. In SEC.~\ref{sec:Ther}, we study static shells 
in a Schwarzschild-AdS system and we explore the parameter space. 
In SEC.~\ref{sec:prob}, we discuss the Euclidean manifold and calculate 
its nucleation probability in the canonical and microcanonical ensemble.
Finally, in SEC.~\ref{sec:con}, we summarize and discuss future topics.

\section{\label{sec:Ther}Static shell in Schwarzschild-AdS}

As we argued in the introduction, thermal activation is described by a periodic solution in the Euclidean manifold. The simplest instanton and the one that exist for all temperatures is a static shell. Before getting into details, let us briefly explain our set up. We start with an initial black hole with mass $M_+$ and we study its decay channels through thermal activation, i.e., static shell instantons, to a lower mass black hole $M_-$ plus a thin-shell in a deeper AdS space. Thus, in this section, we derive the necessary conditions for having a static thin-shell solution and study the parameter range. 

\subsection{Junction conditions}

Let us start by joining two Schwarzschild-AdS spacetimes at a certain radius $r_*$. The ansatz for the metric outside ($+$) and inside ($-$) is given by
\begin{eqnarray}
\label{eq:metric}
ds_{\pm}^{2}= - f_{\pm}(r) dt_\pm^{2} + \frac{1}{f_{\pm}(r)} dr^{2} + r^{2} d\Omega^{2}\,
\end{eqnarray}
where
\begin{eqnarray}
f_{\pm}(r) = 1 - \frac{2GM_{\pm}}{r} + \frac{r^{2}}{\ell_{\pm}^{2}}\,
\end{eqnarray}
and $M_{\pm}$ are the mass parameters of each region, we assume \footnote{In fact, there are solutions for the case $M_->M_+$, if the AdS potential well inside is deep enough. This case resembles that of Ref.~\cite{Garriga:2004nm}, where the cosmological constant decay into a black hole. However, we are interested in the evaporation of black holes, thus we assume $M_+\geq M_-$.} $M_+\geq M_-$ and
$\ell_\pm$ is the AdS length scale. 
These metrics have an horizon at $f_\pm(r_\pm)=0$. Note that the junction is always at $r_*>r_+>r_-$.

Now we place a shell made of matter fields at the junction $r_*$, where the induced metric is given by
\begin{eqnarray}
ds^{2} = - d\tau^{2} + r^{2}_* d\Omega^{2}.
\end{eqnarray}
Continuity of the metric at $r_*$ requires
\begin{eqnarray}
\sqrt{f_{+}(r_*)}\, \frac{dt_+}{d\tau}=\sqrt{f_{-}(r_*)}\, \frac{dt_-}{d\tau}=1\,,
\end{eqnarray}
and the jump in the extrinsic curvature is associated to the presence of the matter field, that is \cite{Israel:1966rt}
\begin{eqnarray}\label{eq:junc}
\epsilon_{-} \sqrt{f_{-}(r_*)} - \epsilon_{+} \sqrt{f_{+}(r_*)} = 4\pi G \sigma r_*,
\end{eqnarray}
where $\epsilon_\pm={\rm sign}({dt_\pm}/{d\tau})$ denote the outward normal direction of the shell and $\sigma$ is the surface energy density of the matter field. We choose that our matter field has an equation of state equals to $-1$, and therefore the conservation of the energy requires $\sigma$ to be a constant. In addition, the fact that the shell is static imposes that Eq.~\eqref{eq:junc} must be satisfied for an infinitesimal displacement away from $r_*$. This leads us to a second condition, that is 
\begin{eqnarray}\label{eq:junc2}
\frac{\epsilon_{-}\,f'_-(r_*)}{\sqrt{f_{-}(r_*)}} -  \frac{\epsilon_{+}\,f'_+(r_*)}{\sqrt{f_{+}(r_*)}}= 8\pi G \sigma,
\end{eqnarray}
where the prime denotes a derivative with respect to $r$, i.e., $f'\equiv \partial f/\partial r$. 

It should be noted that the number of parameters in our set up is six, that is $M_\pm$, $\ell_\pm$, $\sigma$, and $r_*$. However, only three combinations of them are relevant to determine the dynamics of the shell. In practice though, a theoretical model will provide us with $\ell_+$, $\ell_-$, and $\sigma$, so that our free parameters are $M_+$, $M_-$, and $r_*$. After solving Eqs.~\eqref{eq:junc} and \eqref{eq:junc2} we end up with only one free parameter, for example $M_-$ is completely determined by $M_+$ given certain theoretical parameters, i.e., $M_-=M_-(M_+,\ell_\pm,\sigma)$. We devote this subsection to find implicit solutions which although algebraically involved lead to interesting parameter ranges.
For convenience we introduce new dimensionless variables which greatly simplify the analysis. These are given by
\begin{align}
\Delta&\equiv\frac{\ell_+}{8\pi G\sigma}\left(\frac{1}{\ell^2_-}-\frac{1}{\ell_+^2}-\left(4\pi G\sigma\right)^2\right)\quad{\rm ,}\quad
D\equiv\frac{1}{4\pi G\sigma\ell_+}\frac{{M_+}-M_-}{M_+}\,,\\
z_*&\equiv \sqrt{\frac{3}{2}}\frac{r_*}{\ell_+}\,\, {\rm sign}(\Delta)\sqrt{\Delta^2-1}\,,
\end{align}
where $\Delta^2>1$ so that $V(r\to\infty)\to-\infty$. In order to have a clear understanding of these new variables, let us note that $z_*$ might be understood as a rescaled $r_*$. On the other hand, $D$ weights a rescaled relative difference between masses, in our case $D>0$. Note that the sign of $\Delta$ is entirely due to the combination between brackets, which is completely determined by given theoretical parameters. 

With these definitions, we can recast Eqs.~\eqref{eq:junc} and \eqref{eq:junc2}, after squaring twice, in the potential form respectively as
\begin{align}
V(z_*) &= 1  - \frac{c_1}{z_*}-\frac{c_2}{z_*^4}-\frac{2}{3}z_*^2=0\quad {\rm and} \quad 	V'(z_*)=0\,,
\end{align}
where
\begin{eqnarray}
c_1 &=& {\rm sign}(\Delta)\frac{2GM_+}{\ell_+}\sqrt{\frac{3}{2}}\sqrt{\Delta^2-1}\left(1+ \Delta\,D\right),\\
c_2 &=&\frac{9}{4} \left(\frac{GM+}{\ell_+}\right)^2 D^2 \left(\Delta^2-1\right)^{2}.
\end{eqnarray}
In this form, it is clear that the number of parameters that determine the dynamics is three, i.e., $c_1$, $c_2$, and $z_*$. On the other hand, the sign of the extrinsic curvature can be found by squaring Eq.~\eqref{eq:junc} once, which yields
\begin{align}\label{eq:epsilon}
\epsilon_+={\rm sign}\left[D+\Delta\frac{\ell_+}{GM_+}\left(\frac{r_*}{\ell_+}\right)^3\right].
\end{align}

One can find an analytic solution to the position of the shell by solving Eqs.~\eqref{eq:junc} and \eqref{eq:junc2} at the same time, that is
\begin{align}
z_*^{-2}=1+{2}\left({1+{\rm sign}\left(c_1\right)\sqrt{1+\frac{8D^2(\Delta^2-1)}{\left(1+\Delta D\right)^2}}}\right)^{-1}\,,
\end{align}
from which we can readily extract bounds on $z_*$, i.e.,
\begin{eqnarray}
1/2 < &z_*^{2}& < 1 \qquad \,\left(\rm c_1 > 0\right),\\
1 < &z_*^{2}& \qquad \,\,\,\,\,\,\,\,\,\,\,\, \left(\rm c_1 < 0\right).
\end{eqnarray}
Once the position of the shell is known, we can solve for $GM_+/\ell_+$ and $D$ in terms of $z_*$ which yields
\begin{align}\label{eq:A}
\frac{GM_+}{\ell_+}&={\rm sign}\left(\Delta\right)\frac{2\sqrt{2}}{3\sqrt{3}}\frac{z_*}{\sqrt{\Delta^2-1}}\left(1-z_*^2-{z_*}\sqrt{\frac{\Delta^2}{\Delta^2-1}}\sqrt{z_*^2-1/2}\right),\\\label{eq:D}
D&=\frac{2\sqrt{2}}{3\sqrt{3}}\frac{\ell_+}{GM_+}\frac{z_*^2}{\Delta^2-1}\sqrt{z_*^2-1/2}\,.
\end{align}
One may see this system as follows. Given a certain $M_+$, $\ell_+$, $\ell_-$, and $\sigma$, one finds the position of the shell $r_*$ and the mass of the inside black hole $M_-$. In the new variables, given $GM_+/\ell_+$ and $\Delta$, one finds $z_*$ and $D$. Recall that the allowed parameter region must satisfy $\Delta^2>1$. In the following subsection, we further reduce the parameter space.

\begin{figure}
	\begin{center}
		\includegraphics[width=0.49\columnwidth]{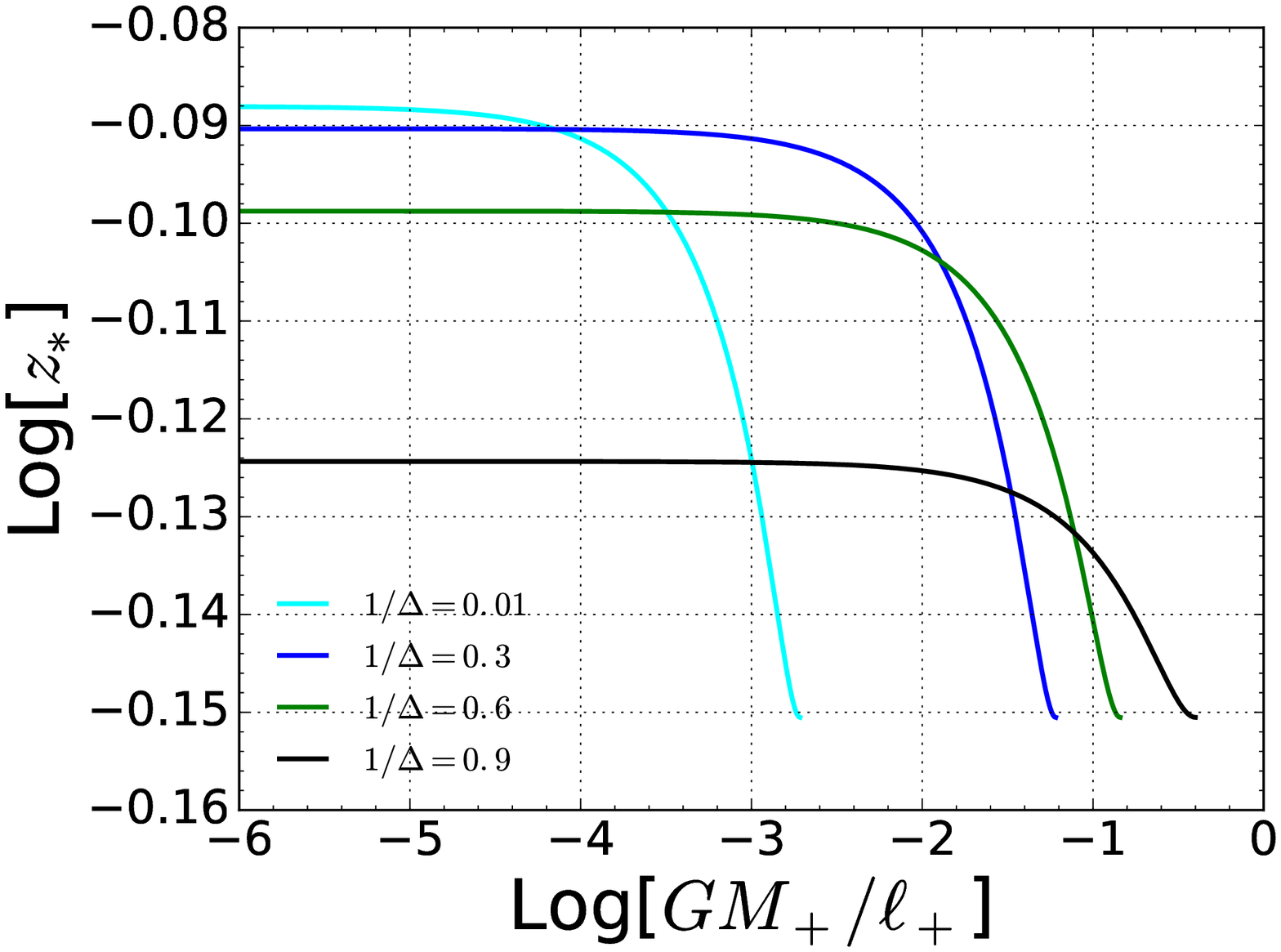}
		\includegraphics[width=0.49\columnwidth]{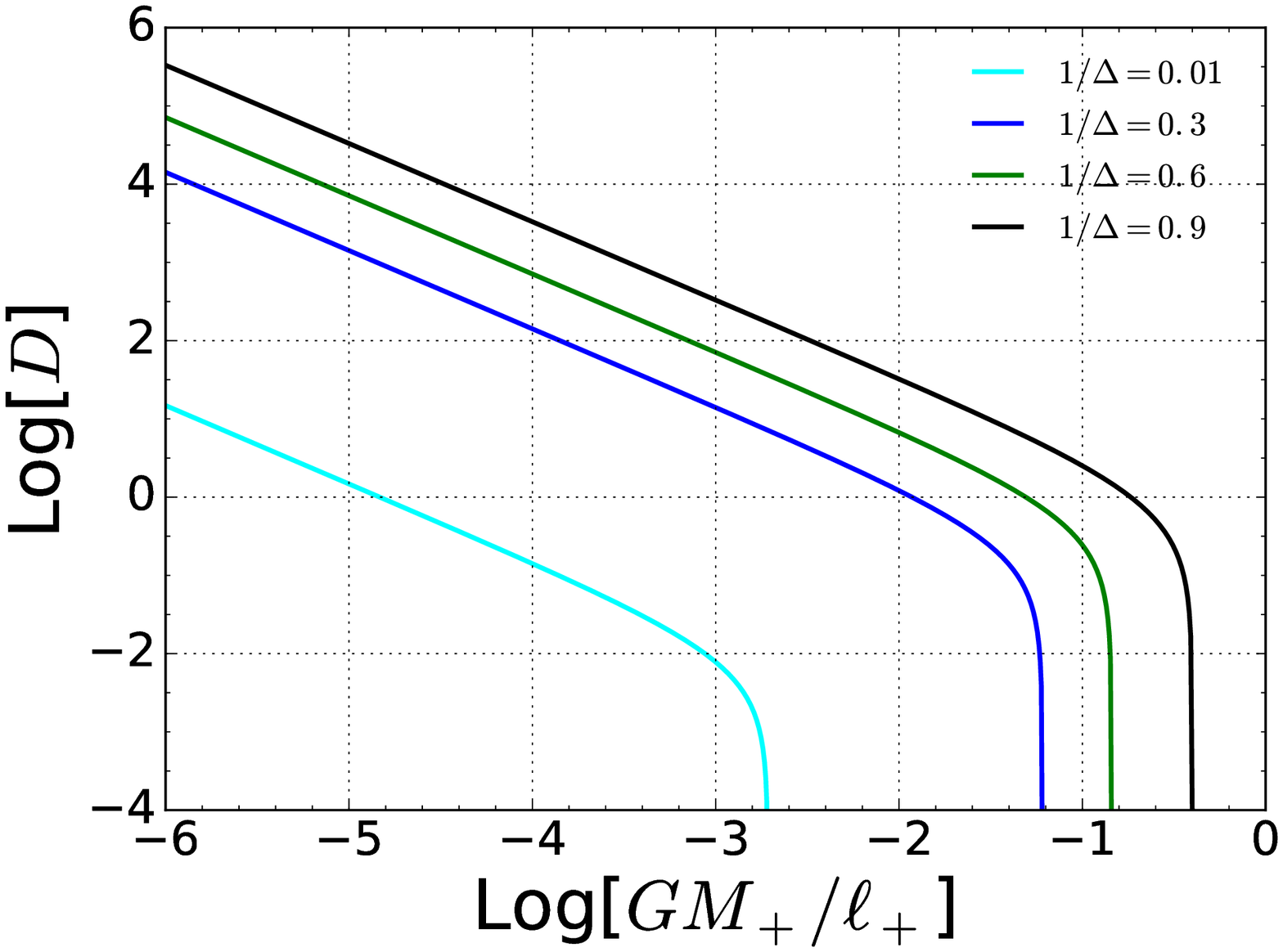}
		\caption{Left: $z_*$ in terms of $GM_+/\ell_+$ after solving Eq.~\eqref{eq:A} for different values of $\Delta$. Note that in general $GM_+/\ell_+<1$ except when $\Delta\sim 1$.. Left: Parameter $D$ in terms of $GM_+/\ell_+$ after solving Eq.~\eqref{eq:D} for different values of $\Delta$. We can se that there is a maximum mass cut-off.}\label{fig:solution1}
	\end{center}
\end{figure}
\begin{figure}
	\begin{center}
		\includegraphics[width=0.45\columnwidth]{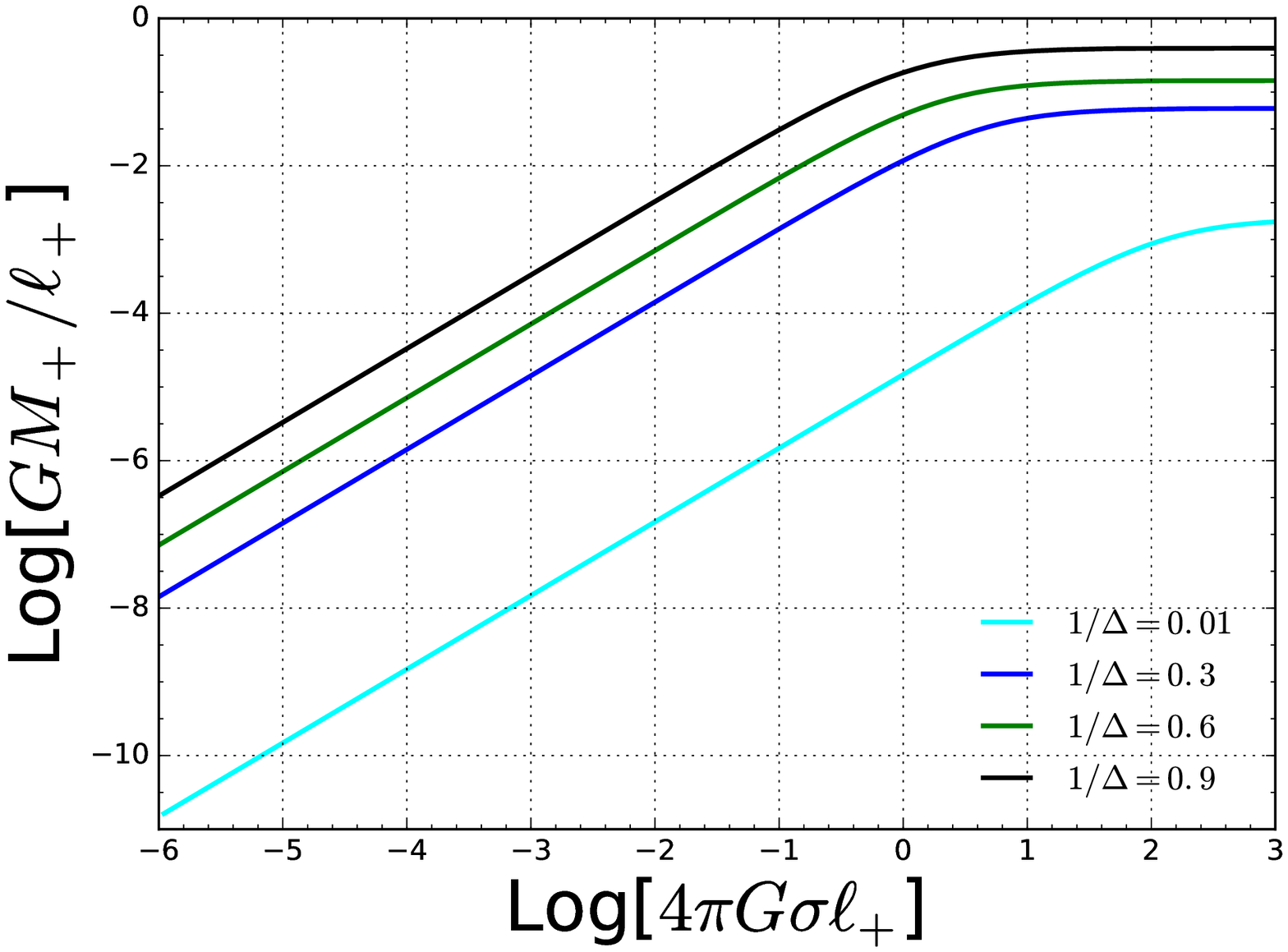}
		\includegraphics[width=0.53\columnwidth]{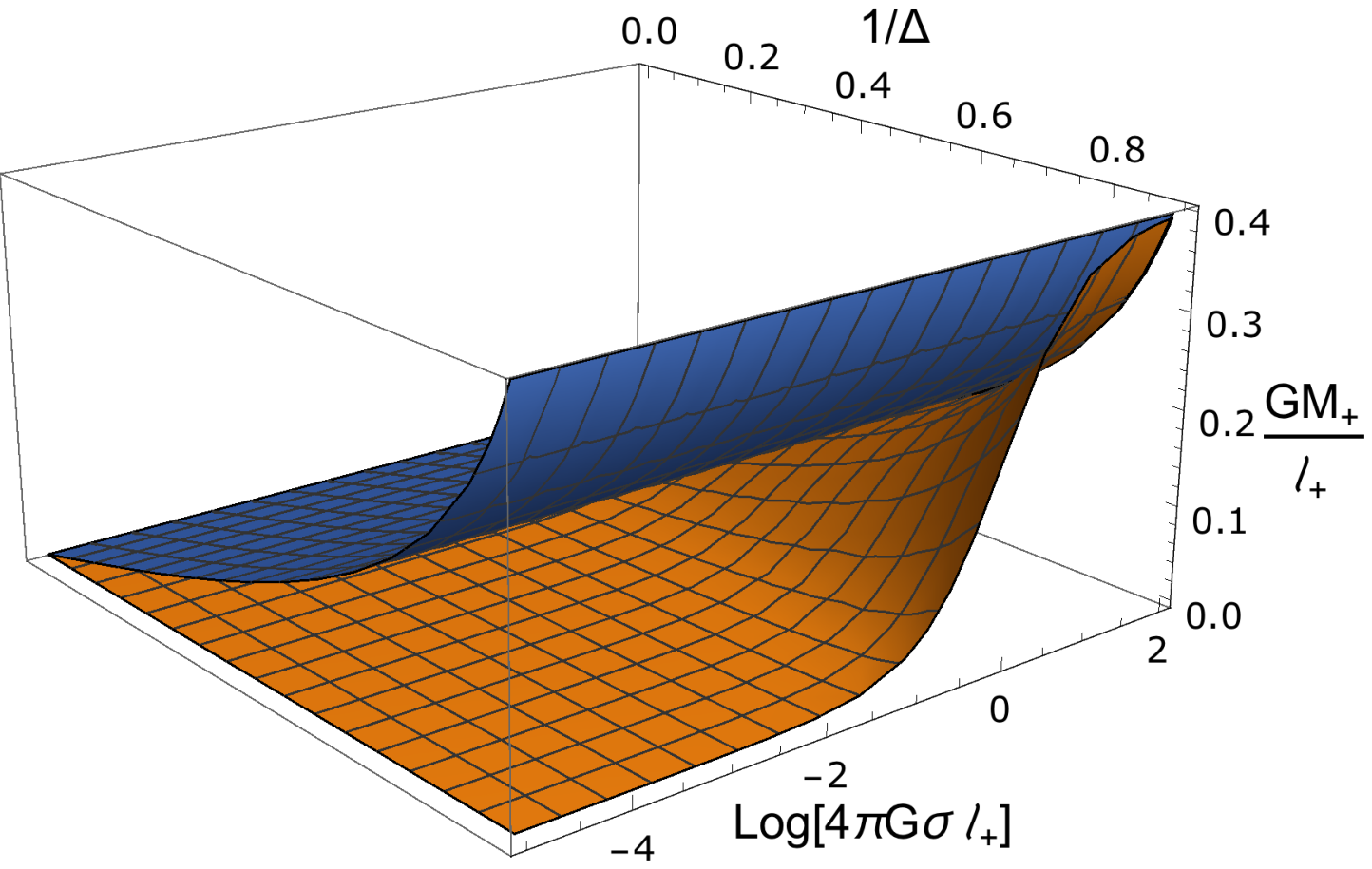}
		\caption{Left: The special case $M_-=0$. It yields a relation between $\sigma$ and $M_+$ in units of $\ell_+$ for a particular value of $\Delta$. Right: Maximum and minimum mass for which the transition is possible in blue and orange respectively. The region in between these two surfaces is the allowed parameter region. Note that they never cross, as it should be.}\label{fig:solution2}
	\end{center}
\end{figure}

\subsection{Parameter space}

\begin{figure}
	\begin{center}
		\includegraphics[scale=0.5]{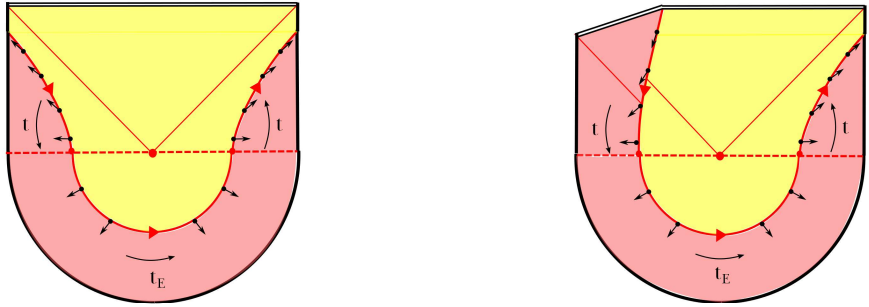}
		\caption{\label{fig:ads_single2} Schematic diagram of the bubble nucleation when the internal mass is non-zero. After the nucleation of the static shell any small perturbation would cause it to either expand or collapse. Left: the shells in both sides expand. Right: the shell in one side collapses while the other expands. We used absorptive AdS boundary conditions for simplicity.}
	\end{center}
\end{figure}

\begin{figure}
	\begin{center}
		\includegraphics[scale=0.5]{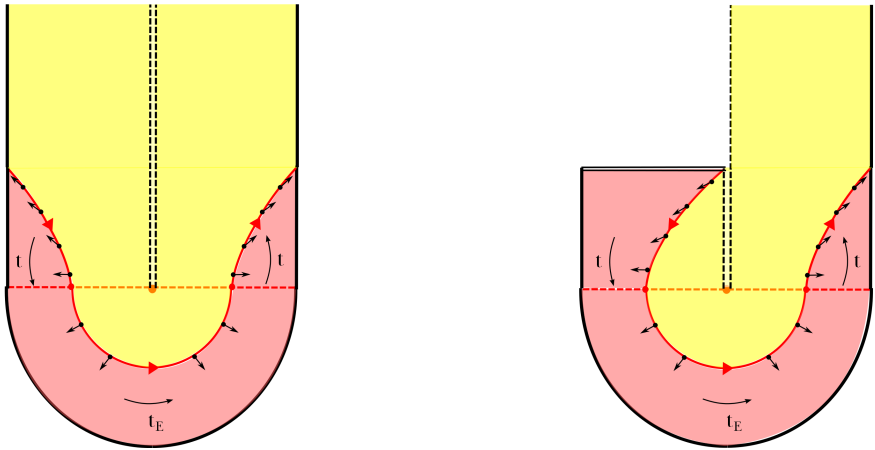}
		\caption{\label{fig:ads_single}Schematic diagram of the bubble nucleation when the internal mass is zero. The dashed line refers to $r=0$ in AdS. We emphasize that after the nucleation the two resulting AdS spaces disconnect. After the nucleation the shell may collapse or expand due to a small perturbation. Left: the shells in both sides expand. Right: the shell in one side collapses while the other expands. We used again absorptive AdS boundary conditions for simplicity.}
	\end{center}
\end{figure}

Now we study further the parameter region. First of all, we are interested in solutions where the outward normal direction does not change sign through the shell, i.e., $\epsilon_+=1$, and therefore the causal structure is given by Figs.~\ref{fig:ads_single2} and \ref{fig:ads_single} (see the next subsection). Interestingly, this excludes the region where $\Delta<0$. Using the solutions for $D$, that is plugging Eq. \eqref{eq:D} into Eq.~\eqref{eq:epsilon}, one gets
\begin{align}
\epsilon_+={\rm sign}\left[\sqrt{z_*^2-1/2}+\Delta\frac{z_*}{\sqrt{\Delta^2-1}}{}\right].
\end{align}
A quick look at the behavior of this function with respect to $z_*$ tells us that $\epsilon_+={\rm sign}\left(\Delta\right)$, which obviously implies $\Delta>0$. Note that the region $c_1<0$ is now excluded. We show numerical solutions for various values of $\Delta$ in Figs.~\ref{fig:solution1} and \ref{fig:solution2}. 

Now that our parameter region has been reduced, we can use the foregoing results to place bounds on the mass of the initial black hole. The minimum of $z_*$, i.e. $z^2_{\rm min}=1/2$, is realized when $D=0$, i.e. $M_{+}=M_{-}$, and yields the maximum of $M_+$ which allows the transition under study. Such maximum mass is determined by a set of theoretical parameters and it is given by
\begin{align}
\left.\frac{GM^{\mathrm{max}}_+}{\ell_+}\right|_{M_+ = M_-} = \frac{1}{3\sqrt{3}}\frac{1}{\sqrt{\Delta^2-1}}\,.
\end{align}
Note that the limit $\Delta\to\infty$ corresponds to $\ell_+\to \infty$ and thus $GM_+^{\mathrm{max}}\to\frac{{8\pi G\sigma}}{3\sqrt{3}}\left({\ell^{-2}_-}-\left(4\pi G\sigma\right)^2\right)$. On the other hand, the minimum value for the outside mass is given by the total evaporation of the black hole, that is $M_-=0$. In that case, we have that
\begin{align}
\left.\frac{GM_+^{\mathrm{min}}}{\ell_+}\right|_{M_-=0}=\frac{2\sqrt{2}}{3\sqrt{3}}\,4\pi G\sigma\ell_+\,\frac{z_{\rm max}^2}{\Delta^2-1}\sqrt{z_{\rm max}^2-1/2}\,,
\end{align}
where
\begin{align}
z_{\rm max}^{-2}
=1+2\left({1+\sqrt{1+8\left(\Delta^2-1\right){\left(4\pi G\sigma\ell_++\Delta\right)^{-2}}}}\right)^{-1}\,
\end{align}
is the maximum value of $z_*$ for a given set of theoretical parameters. Note that as expected, for a fixed theoretical parameters, the lower the mass of the black hole inside the shell the larger the radius of the shell as the difference in energy has to be compensated by increasing the volume of the true vacuum. It is also worth noting that only in the limit \footnote{The exact limit $\Delta=1$ is excluded since the potential would no longer have a maximum at a finite radius.} $\Delta\to1$ the transition is possible for $GM_{+}/\ell_+>1$.

To summarize, there is a static shell solution provided that the initial mass of the black hole lies between $M^{\mathrm{max}}_+ (M_-=M_+)\geq M_+ \geq M_+^{\mathrm{min}}(M_-=0)$ given a set of theoretical parameters that satisfy $\Delta>1$ which implies 
\begin{align}\label{eq:sigma2}
4\pi G \sigma \ell_+< \frac{\ell_+}{\ell_-}-1\,.
\end{align}

We show the allowed parameter range in Fig.~\ref{fig:solution2}. Let us note that this parameter range is also useful in the case of dynamical shells. We can understand the static shell as the limit where the quantum tunneling ceases to exist. Let us call $c_{1,s}$ and $c_{2,s}$ the values of the coefficients of the potential for a static shell, i.e. our solutions to $V(z_*)=V'(z_*)=0$ where $z_*$ is the position of the maximum of the potential. In the general case, quantum tunneling is possible whenever the coefficients $c_1$ and $c_2$ yield $V(z_*)>0$. This achieved for $c_1<c_{1,s}$ and $c_2>c_{2,s}$. Thus, any model in that parameter region presents a dynamical shell with barrier penetration. Note that this may not cover all the possibilities since we imposed $\epsilon_\pm=1$ at $z_*$.

\section{\label{sec:prob}Euclidean manifold and nucleation probability}

We showed that in general there are solutions to a static shell if the initial black hole mass lies between a certain mass range determined by the theoretical parameters. In the Lorentzian signature, this solution corresponds to an unstable static shell at a constant radius. The nucleation configuration is given by analytic continuation of the instanton solution from Euclidean to Lorentzian signatures. A small perturbation at the nucleation will cause the critical shell to either collapse or expand.

\begin{figure}
	\begin{center}\hspace{-1.5cm}
		\includegraphics[width=0.6\columnwidth]{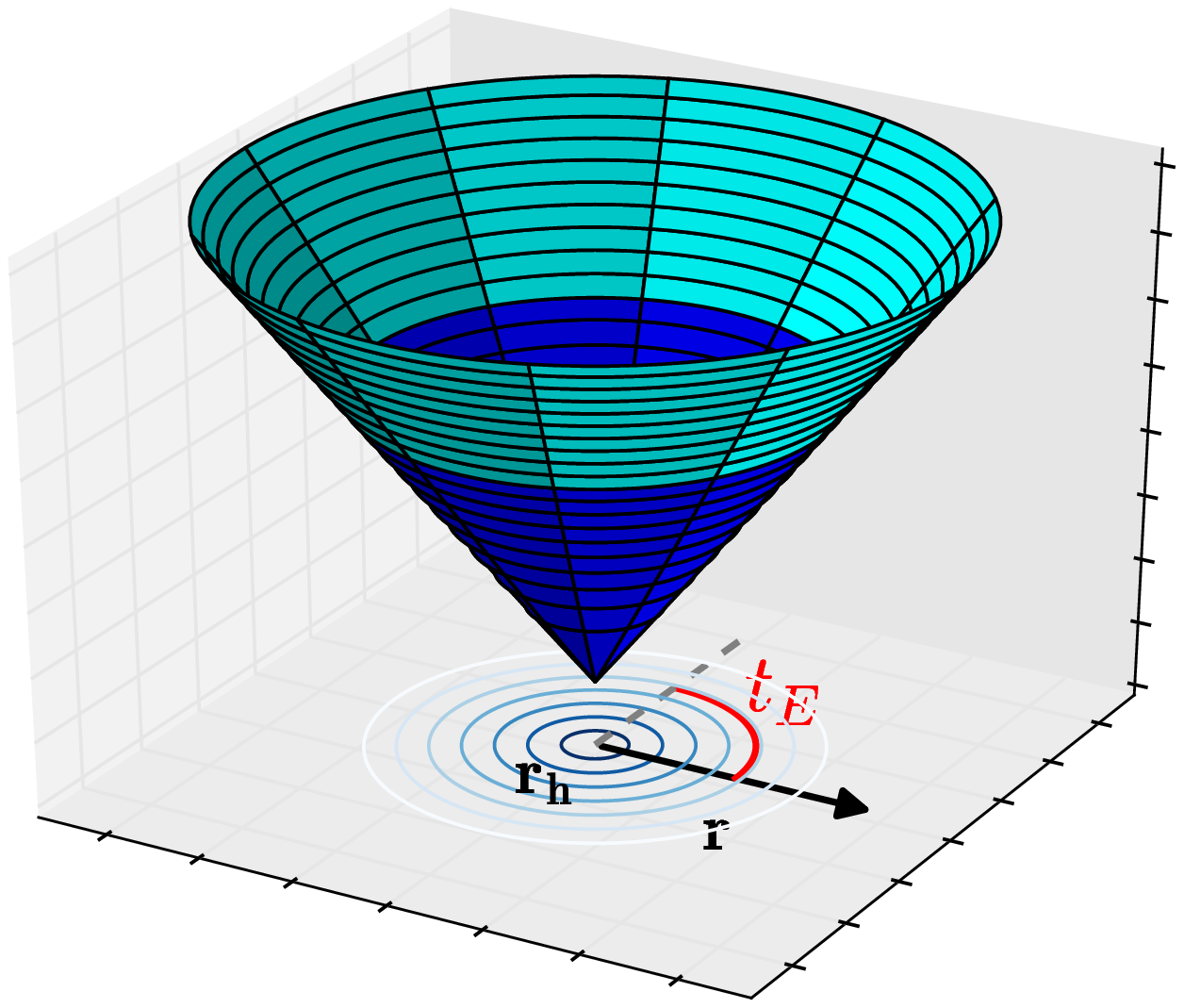}\hspace{-2.7cm}
		\includegraphics[width=0.6\columnwidth]{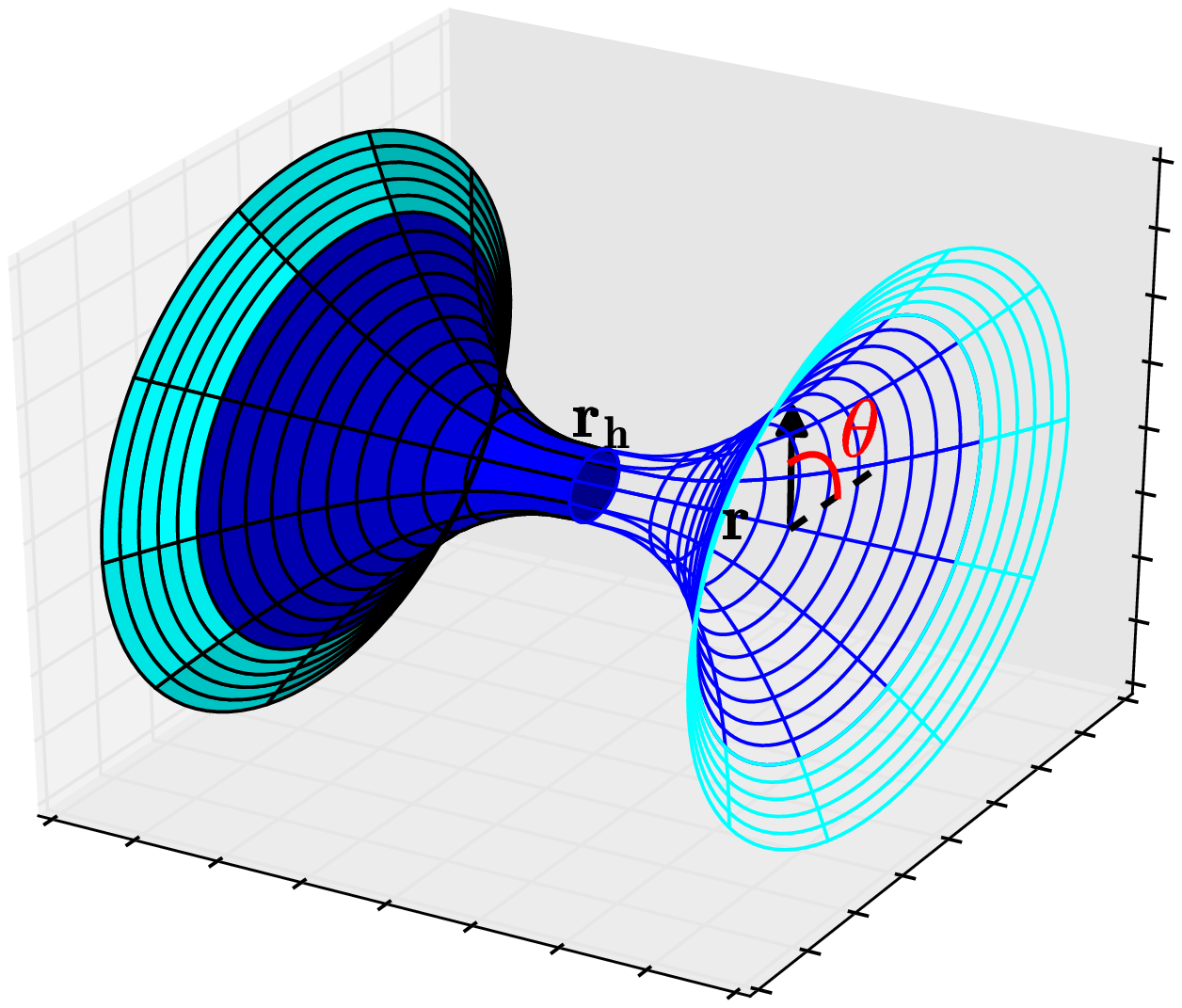}
		\caption{\label{fig:Mnot0}Euclidean manifold for Schwarzschild(-AdS). Light/dark blue respectively refers to outside/inside the shell. Left: Euclidean time vs radius plane embedded in a higher dimension. The cusp can be smoothed out by a suitable choice of euclidean period. Right: Angular vs radius plane with one angular coordinate suppressed. We can see the well-know Schwarzschild throat.}
	\end{center}
\end{figure}

\begin{figure}\hspace{-1.5cm}
	\includegraphics[width=0.6\columnwidth]{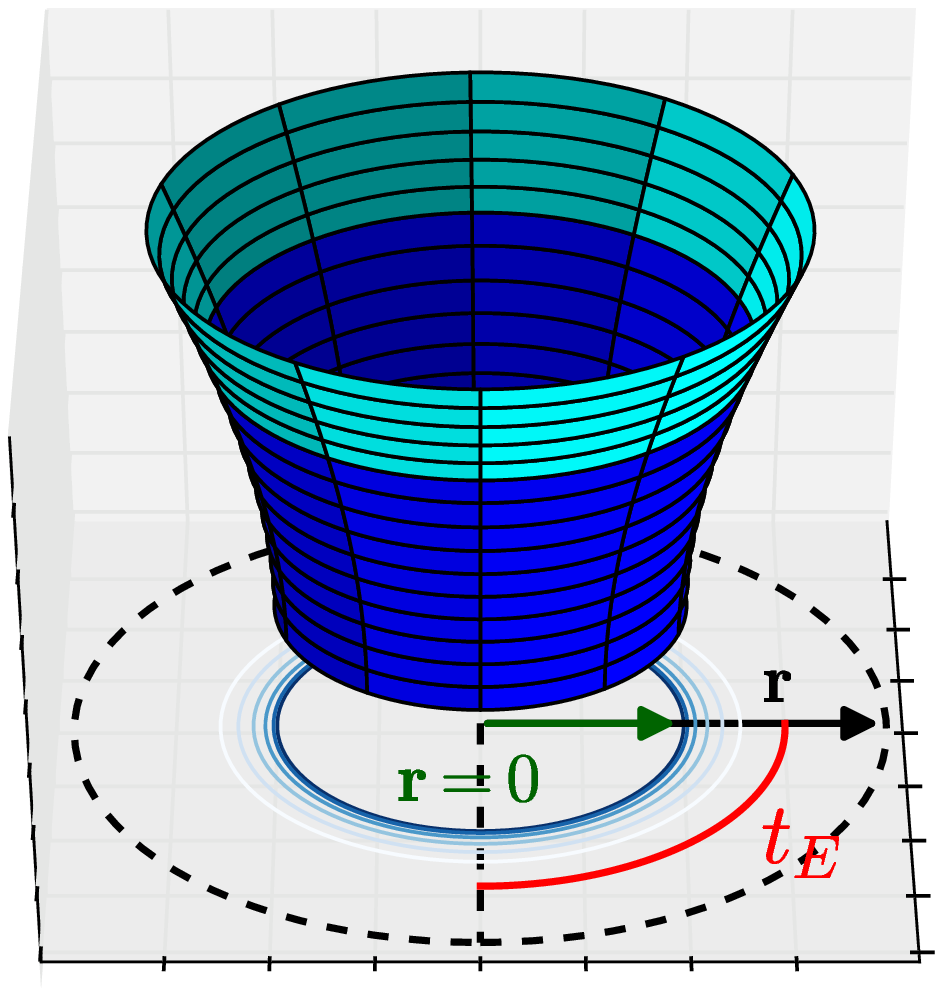}\hspace{-3cm}
	\includegraphics[width=0.6\columnwidth]{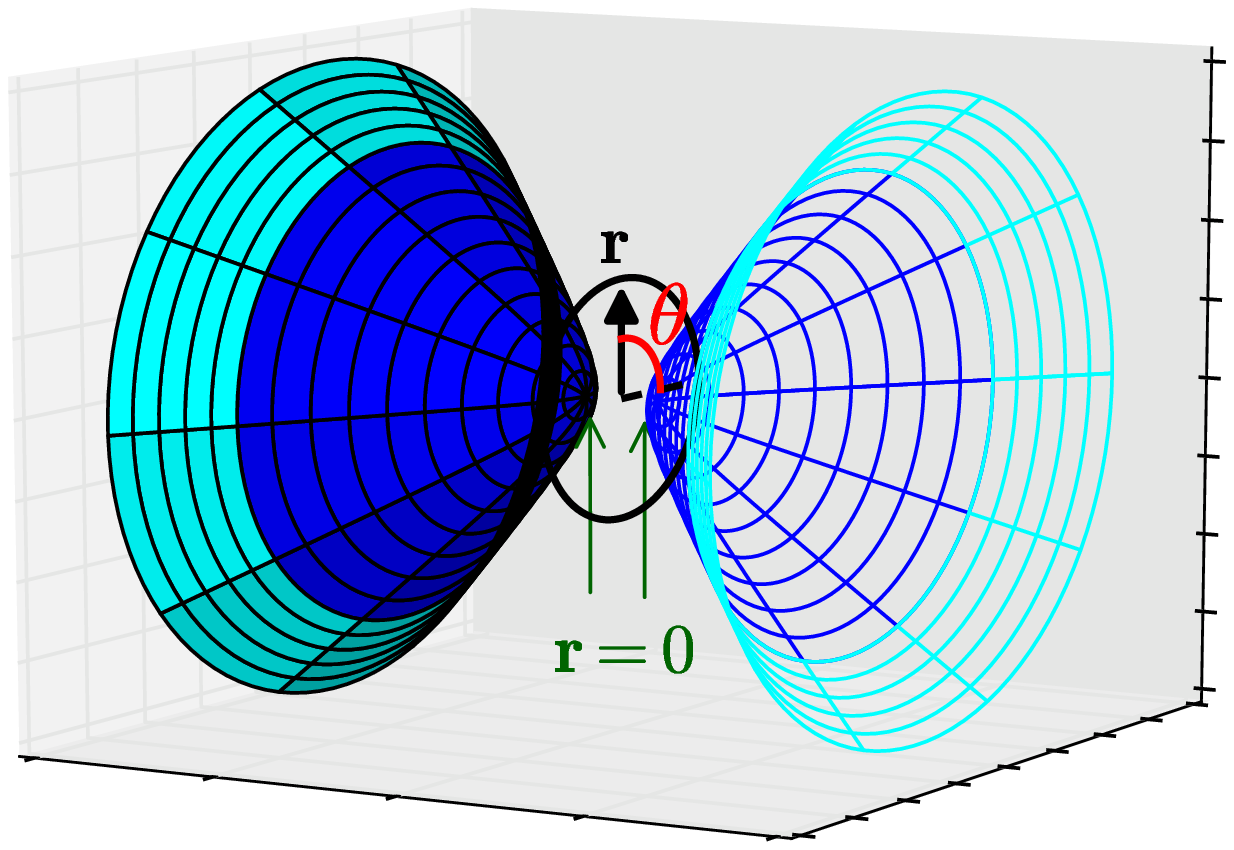}
	\caption{\label{fig:M0}Euclidean manifold for thermal AdS. Light/dark blue respectively refers to outside/inside the shell. Left: Euclidean time vs radius plane embedded in a higher dimension. The manifold reaches $r=0$ which is similar to flat space. Right: Angular vs radius plane with one angular coordinate suppressed. There is no throat connecting both sides but they are connected through the Euclidean time.}
\end{figure}

We can classify two general cases. When $M_{-} > 0$, the internal geometry is a periodically identified Schwarzschild-AdS with a lower mass and deeper AdS (see Figs.~\ref{fig:ads_single2} and \ref{fig:Mnot0}). This might include a conical deficit at the horizon in the Euclidean manifold (see next subsection). A special case corresponds to $M_{-} = 0$, where the internal geometry is a periodically identified AdS (see Fig.\ref{fig:M0}). It is interesting to note that such special case contains a topology change. Now, since the shell is periodic in the Euclidean time, a constant time slice crosses two times the constant radius solution, e.g., the nucleation time slice. This is interpreted as a pair creation of two shells separated by the Einstein-Rosen bridge. In the case where $M_-=0$, the two shells are created in two disconnected pure AdS spaces, most likely due to a pinching of the Schwarzschild throat.

After the nucleation, each shell might either collapse or expand depending on quantum fluctuations of for example a scalar field. If one of them eventually falls into the black hole, then this process is similar to a thin-shell tunneling from inside to outside the event horizon; originally motivated by Hartle and Hawking \cite{Hartle:1976tp} to explain Hawking radiation.


\subsection{Euclidean action and canonical ensemble}

Let us discuss in more detail the Euclidean manifold, the presence of the cusp and the nucleation probability. We start from an initial AdS black hole within the mass that allows a static shell solution. We further assume that the initial black hole is in thermal equilibrium with radiation. Thus, the initial temperature of the system is given by the temperature of the initial black hole. For simplicity, we work under the approximation that radiation makes negligible contribution \cite{Hawking:1982dh}.

After the nucleation the temperature of the bath need not coincide with the temperature of the resulting black hole and therefore, strictly speaking, the system is not in thermodynamical equilibrium. Nevertheless, we can still consider that the system is in equilibrium if the black hole evaporation and absorption time scales are large enough. Let us recall that for radiation they are respectively given by $t_{\mathrm{ev}}\approx10^{71}\left( \frac{M}{M_\odot}\right)^3\,s$ and $t_{\mathrm{abs}}\approx t_{\mathrm{ev}}\left( \frac{T_{\mathrm{BH}}}{T_{\mathrm{rad}}}\right)^4\,s$. Thus, as long as the black hole is large enough, i.e., solar mass, and the temperature of the bath is smaller than that of the black hole, thermodynamical equilibrium is a good approximation. Needless to say, the total evaporation of the black hole, i.e. when $M_-=0$, does not suffer from this problem.

Thermodynamics of a system in thermal equilibrium at a given temperature is described in the canonical ensemble. In thermal QFT, the Helmholtz free energy is given by the Euclidean action. Then, as usual, the nucleation rate is given by
\begin{eqnarray}
\Gamma \propto e^{-B},
\end{eqnarray}
where $B$ is the action difference between the initial and final states, i.e., 
\begin{eqnarray}
B = S_{\mathrm{E}}\left(\mathrm{Final}\right) - S_{\mathrm{E}}\left(\mathrm{Initial}\right)\,,
\end{eqnarray}
where we integrate over a period in the Euclidean manifold. The Euclidean action is given by
\begin{align}
\begin{split}
S_{\mathrm{E}}&=-\frac{1}{16\pi}\int_\pm d^4x\sqrt{g}\left({\cal R}+\frac{6}{\ell_\pm^2}\right)+\sigma\int_{\mathrm{shell}} d^3x \sqrt{h}+\frac{1}{8\pi}\int_{\mathrm{shell}} d^3x \sqrt{h_+}K_+
-\frac{1}{8\pi}\int_{\mathrm{shell}} d^3x \sqrt{h_-}K_-
\,.
\end{split}
\end{align}
Using the equations of motion, integrating and regularizing \cite{Hawking:1982dh,Gregory:2013hja}, i.e., we subtract a periodically identified AdS background with equal period at a fiducial boundary, we get
\begin{align}
S_{\mathrm{E}}=\frac{1}{2}\left(\beta_{\mathrm{in}}\frac{r_*^3}{\ell_-^2}-\beta_{\mathrm{out}}\frac{r_*^3}{\ell_+^2}-\beta_{\mathrm{in}}\frac{r_-^3}{\ell_-^2}-4\pi\beta_*\sigma r_*^2+\beta_{\mathrm{out}}M_+\right)+I_{\mathrm{cusp}}\,,
\end{align}
where $\beta_{\mathrm{in}/\mathrm{out}}$ refer to the period inside and outside of the shell and $\beta_*$ is the period on the shell. Note that if period inside does not match with the temperature of the black hole, then we have to take into account the conical defect. For the moment, let us assume that we need to consider the cusp and show later that it is mandatory. To proceed, recall that the periods are related by the continuity of the metric, i.e.,
\begin{align}\label{eq:betas}
\sqrt{f_-(r_*)}\beta_{\mathrm{in}}=\sqrt{f_+(r_*)}\beta_{\mathrm{out}}=\beta_*\,.
\end{align}
Now, using Eq.~\eqref{eq:junc2} we arrive at
\begin{align}
S_{\mathrm{E}}=-\frac{\beta_{\mathrm{in}}}{2}\left(\frac{r_-^3}{\ell_-^2}+M_-\right)+\beta_{\mathrm{out}}M_++I_{\mathrm{cusp}}\,.
\end{align}



The contribution from the cusp can be easily understood as follows \cite{Gregory:2013hja}. Consider a two dimensional space with an angular deficit, i.e., a cone,
where the line element is given by $ds^2=dr^2+r^2d\phi^2$ with $\phi\to\phi+2\pi\alpha$. 
Then one smooths out the cusp by introducing a small circle of radius $\epsilon$ with a regular geometry inside and take the limit $\epsilon\to0$. This can be modeled by introducing a new function $ds^2=dr^2+A(r)^2d\phi^2$ with a jump in the first derivative, where $\partial_r A(\epsilon)=\alpha$ and $\partial_r A(0)=1$. The Ricci scalar is $R=-2\partial_r^2 A/A$ and therefore after integration we find $\int d^2x \sqrt{-g} R=-4\pi \left(\partial_rA(\epsilon)-\partial_rA(0)\right)=4\pi\left(1-\alpha\right)$. In four-dimensions, we have to add the area of the extra dimensions to the integral which gives
\begin{align}
I_{\mathrm{cusp}}=-\frac{1}{16\pi}\int_{\epsilon} d^4x \sqrt{-g} R=-\pi r_-^2\left(1-\alpha\right)\,.
\end{align}
The factor $\alpha$ is related to the difference in periods, that is
\begin{align}
\alpha={\beta_{\mathrm{in}}}/{\beta_-}\,,
\end{align}
where $\beta_-=4\pi/f'_-(r_-)=4\pi r_-/(1+3r_-^2/\ell_-^2)$ is the temperature associated to the black hole horizon with $M_-$ and $\ell_-$. Plugging this into the action, we find that the free energy is given by
\begin{align}\label{eq:euclideanaction}
S_{\mathrm{E}}=\beta_{+}M_+-\pi r_-^2\,,
\end{align}
where we set $\beta_{\mathrm{out}}=\beta_+$ since the initial state is a black hole in thermal equilibrium. Again $\beta_+=4\pi/f'_+(r_+)$ is the temperature associated to the black hole horizon with parameters $M_+$ and $\ell_+$. This result coincide with the results of reference \cite{Gregory:2013hja}. For the initial system, we should replace $\pi r_-^2$ with $\pi r_+^2$, thus obtaining the well known free energy for AdS black holes \cite{Hawking:1982dh}. It should be noted that in the final state Eq.\eqref{eq:euclideanaction} the temperature $\beta_+$ is fixed by the thermal bath and, therefore, it is independent of the black hole mass.





So far, we have assumed that in general we might encounter a cusp in the Euclidean manifold. Let us show that this is the case. It is easy to see that the condition to have a final smooth manifold is that the period inside the shell coincides with the final black hole temperature, i.e. $\beta_{\rm in}=\beta_-$. For example, such is the case of the Hawking-Page transition. However, they considered that $\ell_-=\ell_+$ which is excluded in our set up since it implies $\Delta<0$.
For this reason, we need to consider the tunneling to a deeper AdS. 

\begin{figure}
	\begin{center}
		\includegraphics[width=0.6\columnwidth]{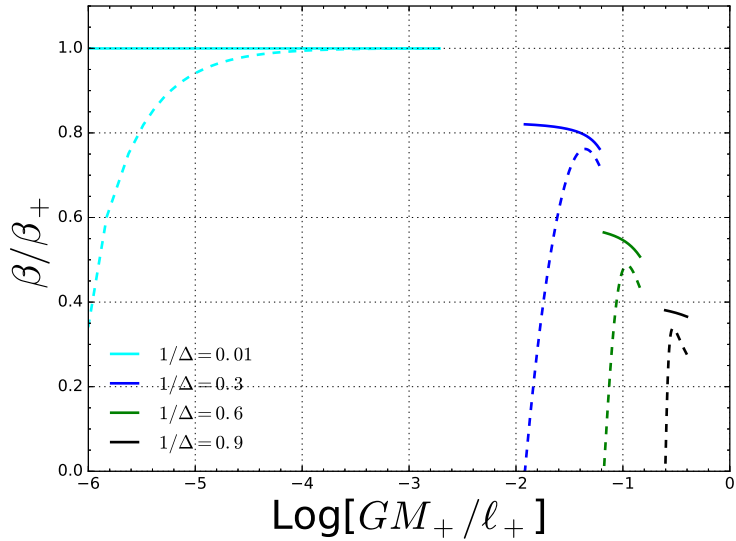}
		\caption{Ratio $\beta_{\rm in}/\beta_{+}$ (continuous lines), where $\beta_{\rm in}$ is the temperature inside the shell, and $\beta_{-}/\beta_{+}$ (dashed lines), where $\beta_{-}$ is the temperature of the inside black hole horizon, as a function of the initial black hole mass. We plotted it for $\ell_+=3\ell_-$ and we let $\Delta$, i.e. $4\pi G\sigma\ell_+$, vary. Notice that the lines never intersect, therefore $\beta_{in}\neq\beta_{-}$ in general. For the light blue line, i.e. $\Delta=100$, one check that ${\beta_{in}-\beta_{-}}>10^{-5}\beta_{+}$. Obviously, $\beta_-=0$ when $M_-=0$.}\label{fig:betas}
	\end{center}
\end{figure}

It is easy to guess that the condition $\beta_{\rm in}=\beta_-$ might not be fulfilled in general from the following. Take Eq.~ \eqref{eq:betas} with $\beta_{\rm out}=\beta_+$ and $\beta_{\rm in}=\beta_-$. Let us remind the reader that $\beta_{\rm out}=\beta_+$ is fixed by the initial thermal equilibrium. Then the question is whether the condition on $\beta_{\rm in}$ is compatible with the parameters determined by the junction conditions, that is
\begin{align}\label{eq:eqtemp}
\frac{\beta_{\rm in}}{\beta_{+}}=\sqrt{\frac{f_+(r_*)}{f_-(r_*)}}\stackrel{?}{=}\frac{\beta_{-}}{\beta_{+}}=\frac{f'_+(r_+)}{f'_-(r_-)}\,.
\end{align}
where we used the definition of temperature in the last step. Note the behavior of the left hand side in Eq.\eqref{eq:eqtemp} is completely different from that in the right hand side. Thus, the equality could be satisfied only by a very particular set of parameters. In fact, we numerically show in Fig.~\ref{fig:betas} that the $\beta_{\rm in}=\beta_-$ condition is not satisfied at all for a wide range of parameters. Let us note that the only parameter we fixed in Fig.~\ref{fig:betas} is the ratio between AdS lenght scales, i.e. $\ell_+/\ell_-$. Varying this ratio does not seem to affect the general behavior of the curves. Therefore, we must consider the presence of a cusp if one wants to allow this decay channel. 

\begin{figure}
	\begin{center}
		\includegraphics[width=0.49\columnwidth]{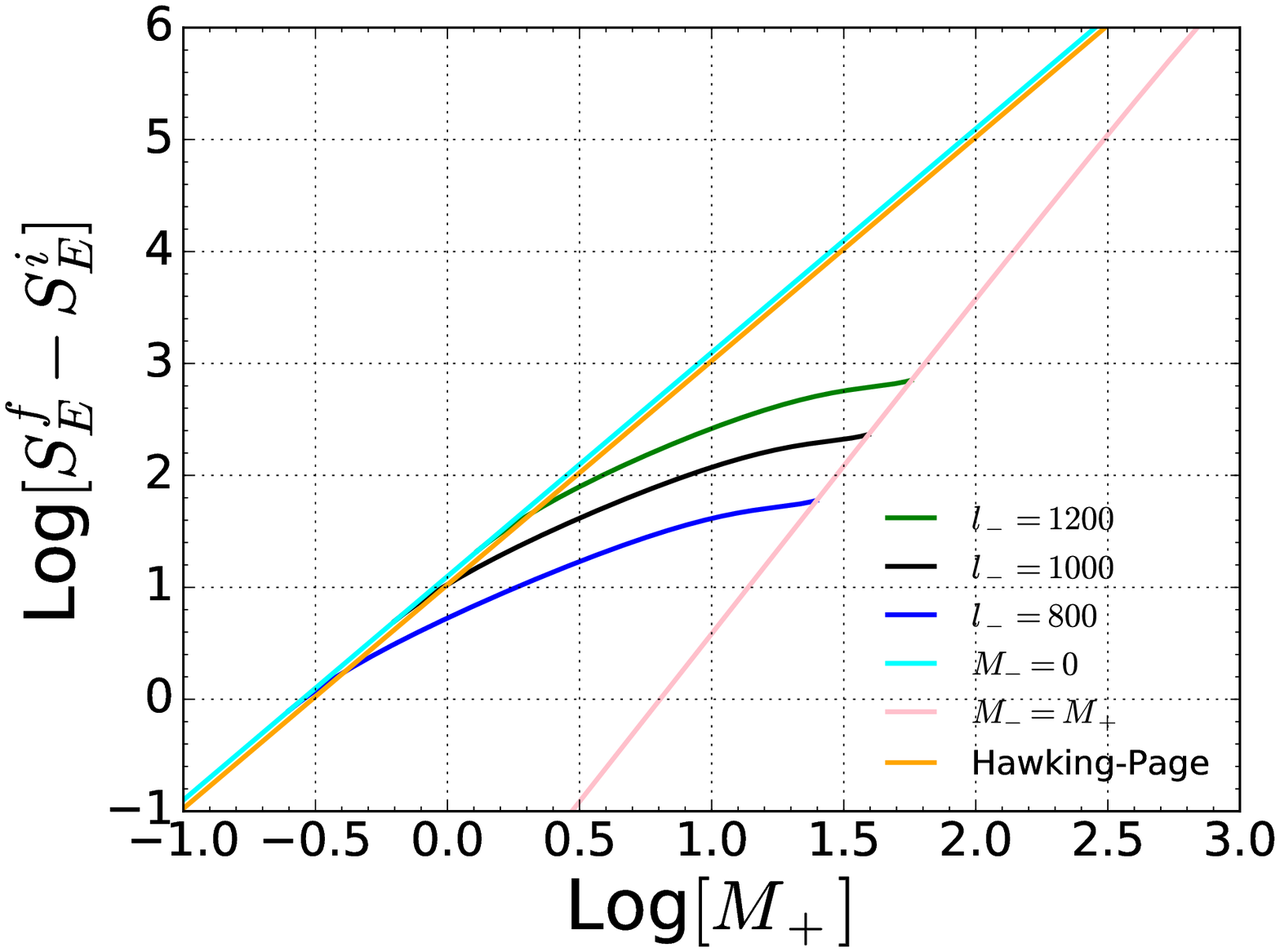}
		\includegraphics[width=0.49\columnwidth]{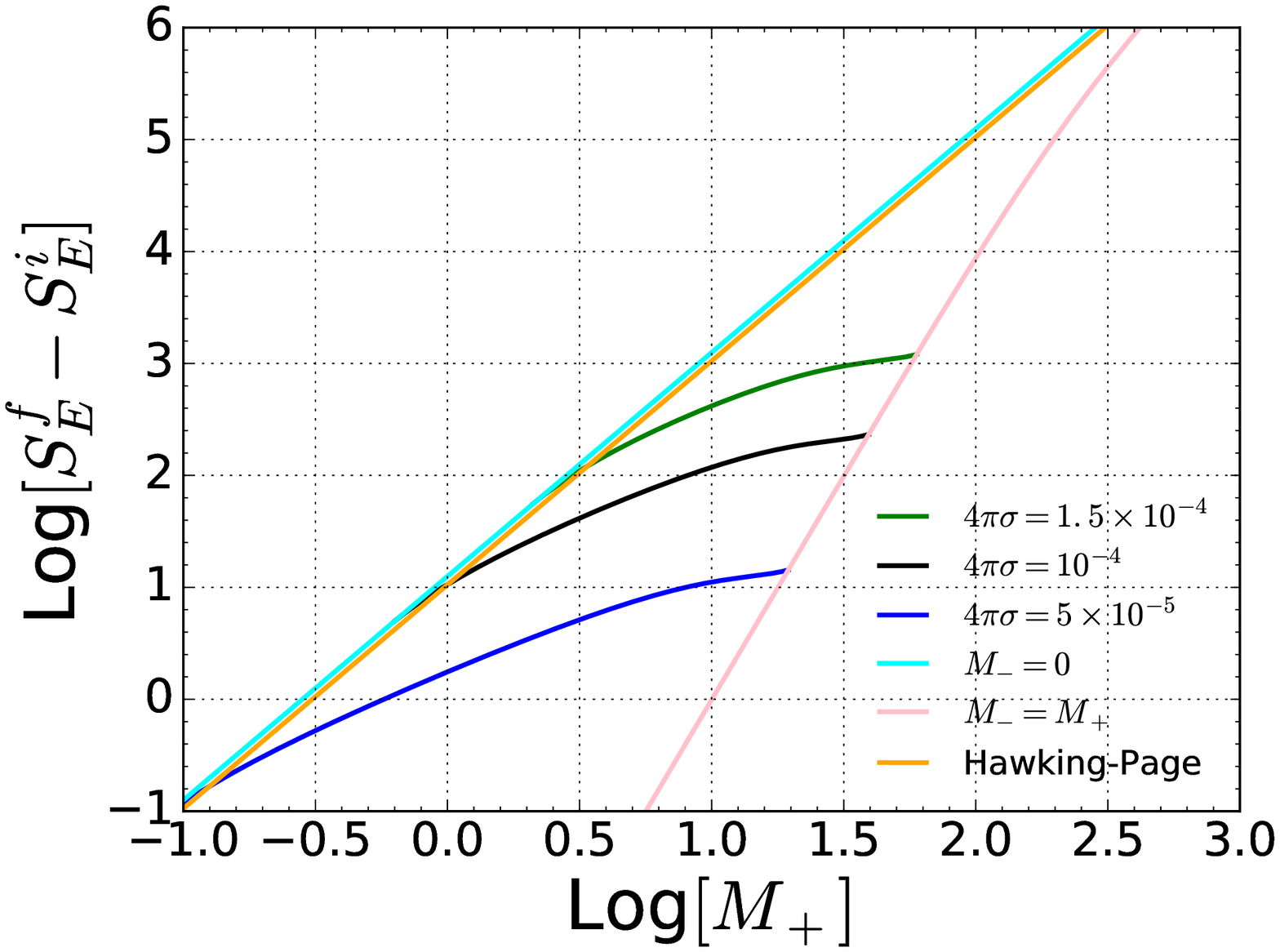}
		\caption{\label{fig:action}Difference of action in units of $G=1$ for $\ell_+=10000$ as a function of $M_+$. On the left we plot various inside AdS lenght scale, i.e. $\ell_-=\,800\,,1000\,,1200$, and on the right we change the surface energy density, that is $4\pi\sigma=\,(15\,,10\,,5)\times10^{-5}$. In green and pink we show the upper and lower bound for the outside mass respectively. The probability for the Hawking page phase transition, i.e. the black hole decay to radiation, is plotted in orange. Note that all parameters can be rescaled with $M$ and then the action must be multiplied by a factor $4 M^2$. }
	\end{center}
\end{figure}

As a result, using Eq.~\eqref{eq:euclideanaction}, the tunneling probability in the canonical ensemble is given by
\begin{align}
S_{\mathrm{E}}^f-S_{\mathrm{E}}^i=\frac{1}{4G}\left(A_+-A_-\right)\,,
\end{align}
which only depends on the difference of areas of the initial and final black holes (this is also consistent with a diffent approach \cite{Farhi:1989yr,Chen:2015ibc}). We show in Fig.~\ref{fig:action} the behavior of the tunneling probability in terms of the initial mass of the black hole in units of $G=1$. The allowed region falls between the maximum mass (lower pink line) and minimum mass (upper green line). As expected, the complete evaporation is the most suppressed channel. For completeness, we compare our results with the probability of the Hawking-Page phase transition to radiation. Note that in terms of probability this two processes are similar.

\subsection{Microcanonical ensemble}

In practice, thermal equilibrium for a black hole system is an extremely long state to achieve. A more realistic set up is to consider transitions with a fixed energy. Therefore, we need to consider the \textit{microcanonical ensemble}. It is known that the canonical and the microcanonical ensembles are related by an inverse Laplace transformation \cite{Hawking:1982dh}, that is
\begin{align}
N(E)= \frac{1}{2\pi i}\int_{-i\infty}^{i\infty} {\cal Z}(\beta){\rm e}^{\beta E} d\beta=\frac{1}{2\pi i}\int_{-i\infty}^{i\infty} \exp\left(\beta E-S_E(\beta)\right) d\beta\,.
\end{align}
Where ${\cal Z}(\beta)$ and $N(E)$ are the canonical and microcanonical partition functions respectively. In the usual case, the temperature of the system depends on the mass of the black hole and the integral is calculated using the steepest-descent approximation \cite{Hawking:1982dh}. However, recall that on our case, the temperature in the final state is fixed by the thermal bath and therefore it is completely unrelated to the mass of the final black hole. Thus if we assume that radiation does not backreact on the geometry we have that
\begin{align}
N(E)= \frac{{\rm e}^{\pi r_-^2}}{2\pi}\int_{-\infty}^{\infty}{\rm e}^{\beta \left(E-M_+\right)} d\beta={\rm e}^{\pi r_-^2}\delta\left(E-M_+\right)\,,
\end{align}
where we used that $S_{\mathrm{E}}(\beta)=\beta GM_+- \pi r_-^2$ and that $M_+$ and $r_-$ do not depend on the temperature. That is, the number of states of the system is proportional to the area of the black hole. Therefore, transition to a lower mass black hole is suppressed by the difference in the number of states, i.e., it is suppressed by exponential of the difference in entropies. This results coincides with the canonical ensemble and gives more support to consider such transition.

\section{\label{sec:con}Conclusions}

We have considered thermal activation around AdS black holes. The nucleation of thin-shell bubbles out of a thermal bath is mediated by a static shell instanton and therefore trivially periodic in the Euclidean time. We build static shell solutions and we studied their parameter space. We have found that for a given set of theoretical parameters, that is the vacuum and shell energy densities, there is a static shell solution if the initial black hole lies within a certain mass range determined by the theoretical parameters. In particular, if there exists a solution for a given theoretical parameters then there is always the solution where the black hole completely evaporates.

Later, we assumed that the periodicity in the Euclidean time is given by a thermal bath, fixed by the thermal equilibrium between the initial black hole and radiation. We have shown that in general by fixing the initial temperature, we had to consider the presence of a conical deficit at the horizon due to the lack of solution with equal temperature. The presence of such a cusp can be dealt with a smoothing regularization scheme and the Euclidean action simply coincides with the usual result. However, let us stress that the crucial difference is that the temperature of the final system is not related to the final black hole mass. Then, we have discussed the nucleation probability in the canonical and microcanonical ensembles and we have shown that it depends on the difference of areas from the initial to the final black hole. The case of the complete evaporation of the black hole has a probability very close to the Hawking-Page phase transition.

After a pair nucleation of shells over the Einstein-Rosen bridge, each shell will either collapse or expand most likely due to quantum fluctuations along the Euclidean manifold. Will the two shells be entangled? In addition, this looks similar to a particle tunneling from inside to outside the horizon. Can this be further generalized to various thermal phenomena around the black hole? We leave them for future work.

Let us finish by pointing out some important remarks. First, as mentioned in the introduction the fact that we start with a system in thermal equilibrium might obscure its relation with the information loss paradox. However, as we have shown there exists in general solutions for which the black hole completely evaporates. How the shell carries the information or how the information is recovered is an interesting point which is far from the scope of this paper. In addition to this, a realistic system would not be perfectly thermalized (hence, information would be still there), but it could be well approximated by a quasi-thermalized system. Then what will be the implication of these thermal activations? This would be an interesting question for a future work.

Second, in a more realistic set up, the cusp at the horizon should be regularized by some other field around the black hole to account for the angular deficit, e.g., a string. This regularization should not affect the main results of this paper, which is that the black hole could evaporate completely. We leave for future work how this instanton mediates a transition where there is a conical deficit (or a topological change for the case of a complete evaporation) in the final Euclidean manifold.

\section*{Acknowledgments}
DY was supported by Leung Center for Cosmology and Particle Astrophysics (LeCosPA) of
National Taiwan University (103R4000). This work was supported in part by the MEXT KAKENHI Nos.~15H05888 and 15K21733. G.D. would like to thank S. Ansoldi, J. Garriga, T. Tanaka and S. Yokoyama for fruitful discussions and comments on thermal activation.

\end{document}